# The Interactive Minority Game: a Web-based investigation of human market interactions

Paolo Laureti, Peter Ruch, Joseph Wakeling[*] and Yi-Cheng Zhang

*Département de Physique, Université de Fribourg, Pérolles, CH-1700 Fribourg, Switzerland*

(16 July 2003)

**Abstract**

The unprecedented access offered by the World Wide Web brings with it the potential to gather huge amounts of data on human activities. Here we exploit this by using a toy model of financial markets, the Minority Game (MG), to investigate human speculative trading behaviour and information capacity. Hundreds of individuals have played a total of tens of thousands of game turns against computer-controlled agents in the Web-based *Interactive Minority Game*. The analytical understanding of the MG permits fine-tuning of the market situations encountered, allowing for investigation of human behaviour in a variety of controlled environments. In particular, our results indicate a transition in players' decision-making, as the markets become more difficult, between deductive behaviour making use of short-term trends in the market, and highly repetitive behaviour that ignores entirely the market history, yet outperforms random decision-making.

*PACS:* 02.50.Le; 89.65.Gh; 89.70.+c

*Keywords:* Decision theory and game theory; Economics and financial markets; Information theory; Internet experiments

Experimental games and their theoretical offspring have been a fruitful research direction for various disciplines, particularly psychology [1-6] and economics [7-11], but elsewhere as well [12-14]. The advantage of this approach is that the simplified game environment allows for controlled investigation of human behaviour while still potentially maintaining the essential features of real-world situations. A notable example in recent years has been the so-called "market entry" games [9,11], where traders must decide whether or not to join a market based on knowledge of its capacity and of their competitors' past actions. These games have generated much interest as examples of situations where the insight provided by classical economic theory is limited, and an experimental approach was thought essential [11,15,16].

---

[*] Corresponding author. Email: joseph.wakeling@unifr.ch.

By coincidence, a theoretical approach has been developed by the statistical physics community for an independently created game that has many similarities to the market entry class: the Minority Game (MG) [17]. Economic agents are endowed with simple strategies and learn inductively, as suggested by Arthur [18]. A rich market dynamics emerges, whose properties depend on only a few simple parameters [19-21]. These results have recently led some authors to return to more traditional experiments, playing the MG with small groups of humans [22,23].

Our approach here has instead been to make use of the understanding of the theoretical game, by having individual humans play against computer-controlled "MG agents". We can thus fine-tune the market situation the player encounters, and provide a variety of controlled environments in which to investigate human behaviour. Because we only ever engage individual players, we have been able to make use of the immense access provided by the World Wide Web[1], presenting the game via an online interface. Since being launched a year ago [25], hundreds of players have played a total of tens of thousands of game turns in the *Interactive Minority Game*.

The player is presented with a "price" history of the past 50 time steps of a market (Fig. 1) in which he is one of $N$ traders, the others being MG agents. At each time step $t$, each individual $i$ must choose independently, based on the market history, between two actions, $a_i(t) = +1$ or $-1$ (say, "buy" or "sell")[2]. They then receive points given by the formula,

$$g_i(t) = -a_i(t)A(t), \qquad (1)$$

where $A(t) = \sum_{i=1}^{N} a_i(t)$ is the aggregate action of the population at time $t$. Thus, those whose choice is in the minority gain $|A(t)|$ points, and those in the majority lose this amount; a player's average gain per turn, $\langle g_i(t) \rangle$, can be taken as a measure of his success. (Note that the Minority Game is a non-zero sum game.) From the values $A(t)$ we can generate the price history[3] shown to the human player [26,27],

$$P(t+1) = P(t) + A(t). \qquad (2)$$

Five different markets are available to choose from, whose names give a rough order of increasing difficulty. Broadly these can be divided into two groups, determined by a control parameter $\alpha = 2^M/N$, where $M$ is the MG agents' game memory. (Table 1 gives the statistics for the different levels.) The first three markets are in a "symmetric" phase with large price fluctuations, informationally efficient if looked at with memory $M$ or less, yet inefficient if looked at with a longer memory value, with noticeable herding effects among MG agents [19]—*Easy*, *Apprentice* and *Trader*, with average fluctuations decreasing respectively. The other two levels, *Professional* and *Guru*, are in an "asymmetric" phase with fluctuations less than or equal to those which would result from players making random decisions—yet with small amounts of usable information still present in the price history if viewed with memory *M*.

---

[1] For instance, the SETI@home project (http://setiathome.ssl.berkeley.edu/) was able to gain the processing power of over a million computers by requesting net users to download a screensaver that also acted as a data analysis program [24].

[2] For ease of play, the human player is in fact asked to predict the direction of the next market price change (i.e., the next majority action); his own action (the opposite) is inferred from this. For more details on the MG, including details of computer-controlled agents' decision-making process, see Refs. 17, 20 and 25.

[3] Technically $P(t)$ is the logarithm of the true market price, but presenting the market history in this form communicates far better what is actually occurring in the Minority Game being played.

Each market consists of $N = 95$ players, including the human. All were started with a 1000-turn introductory period with computer-controlled agents only; the subsequent interactive games are continuous, with each player picking up the price history where the last leaves it. Thus, a continuous price history (Eq. 2) of each market can be generated[4].

As a first step to analysing human performance, one can consider the human player's rank in the game compared to the computer-controlled agents (Fig. 2). A marked difference can be observed depending on the market phase. In the symmetric-phase markets humans almost always gain rank 1, clearly exploiting information the MG agents cannot see, while in the asymmetric markets humans do much worse, with most players being located in the lower half of the ranks (though their performance spans the complete range of possibilities)—a much fairer game.

An alternative point of view is given by considering humans' average gain per turn in the game, as compared to average gain per turn for the regular MG agents and other types of computer-controlled player. One can use the distribution of scores as a function of rank among other humans, which we display for Easy and Apprentice levels (Fig. 3a), or (to better compare to computer-controlled players) the cumulative distribution of scores, which we display for Guru level (Fig. 3b). In the former case humans can easily exploit the herding effects in the market, doing much better than most MG agents and usually better than even the best MG agent. The latter, fairer market proves much more difficult: human performance covers a range from the very lowest to the very highest scores achieved by MG agents, but the MG agents generally do better (reflecting the results of Fig. 2). In all cases, however, humans do consistently better than random decision-making ("noise trading").

Insight into human decision-making processes can be gained by information-theoretic analysis. If we denote by $a_*(t)$ the human's action at time *t*, and $\mu_m(t)$ the market history of length *m* preceding this decision, then predictability of human action can be measured by the *information entropy* [28], $H(a_* | \mu_m)$, of the sequence $a_*(t)$ conditional on $\mu_m(t)$. More conveniently, we can use the *information gain ratio*[5] defined by

$$I_m := 1 - \frac{H(a_* | \mu_m)}{H(a_*)}, \qquad (3)$$

---

[4] Games are only included in this history if they last more than 100 turns. The computer-controlled agents are also inherited from game to game, with evolutionary rules built in to allow them to adapt to each individual player and not become "frozen" into rigid responses.

[5] This measure of association between *m*-bit strings and subsequent actions is sometimes also referred to as the *uncertainty coefficient*. A similar analysis of actions in response to histories has independently been proposed in Ref. 22.

which gives us extremal values $I_m = 1$ and 0, meaning respectively that human action is completely predictable or completely random with respect to market histories of length $m$.

Fig. 4 shows how $I_m$ changes with $m$ at different levels, both over the entire continuous market history (left) and for handpicked players with high scores (right) who had each played around 1000 turns. When computer-controlled agents have shorter memory values $I_m$ (the Easy and Apprentice levels), all humans are able to spot these values and exploit resulting patterns, playing as if they were basing most of their decisions on market histories of length $M + 1$. This is particularly noticeable for the handpicked players, for whom the discontinuities in $I_m$ are very large at Easy and Apprentice levels, and marked even at Trader—indicating considerable exploitation of patterns in the market.

In the more difficult levels, as agents' memory is increased, no clear discontinuities in $I_m$ can be observed. Instead players tend to ignore the market history entirely and simply repeat their immediately preceding action with large probability (~0.8 at Professional and Guru). This repetitive behaviour is even stronger among the handpicked players. Further analysis shows that while human decisions in the easier markets are correlated with the long-term trend of the market, this correlation decreases as the markets become more difficult, being close to zero at Professional and Guru levels. Thus, at these levels players appear to be ignoring all aspects of the information presented by the market. This behaviour in the more difficult markets reflects the observations of other authors on humans playing the Minority Game in groups [*22,23*], but in our case the tendency to repetition is even stronger.

A final point of view is provided by examining the human's market impact, by considering how volatility and the average gain per turn of computer-controlled agents change as the human enters and leaves. The former can be measured by the normalized variance of the market fluctuations:

$$\frac{\sigma^2}{N} = \frac{\langle A^2(t) \rangle}{N}. \qquad (4)$$

In the easier markets a good player can consistently decrease $\sigma^2/N$, most obviously at Easy level (Fig. 5, dotted line) but also observable at Apprentice. This has the unintended benefit of actually *increasing* the gain of the computer-controlled players (Fig. 5, solid line): thus, a symbiosis exists between the "selfish" human and the MG agents, who gain from a speculator decreasing volatility by exploiting market inefficiencies they cannot themselves observe. So to speak, the human is "doing good by doing well". By contrast at Trader, Professional and Guru levels the effect of the human player is too small to make any clear statement. Since volatility at Trader level is still greater-than-random, this suggests that there is a limit to untrained humans' ability to arbitrage: when volatility is below a certain level, there is not much they can do to decrease it further, despite inefficiencies still being present.

In summary, whereas most studies of market entry games have concentrated in the main on whether players can coordinate to an equilibrium [9,11,18], the different controlled environments provided by the analytically understood Minority Game allow for quantitative measurement of important factors in economic decision-making. The differences in performance between the levels are likely to be in part a result of the different memory values of computer-controlled agents, suggesting that humans may have a maximum length of market history over which they are able to consider patterns. However, emphasis must also be put on the smaller fluctuations in the more difficult markets, which mean that short-term patterns cannot survive for long under market impact.

The transition we observe, based on the market phase, between behaviour utilizing short-term patterns in the market and the long-term direction of price movement, and simple repetitive behaviour that ignores all aspects of the market history, provides a confirmation of Arthur's suggestion [18] that beyond a certain level of complexity human logical capacity can no longer cope. When observed elsewhere in human economic decision-making, repetitive behaviour has been criticized as "illogical" [22], as it potentially provides information that competitors could exploit. Yet, as we have seen here, this behaviour actually consistently outperforms random actions, which provide no such information. This might suggest that the transition is not between learning and ignorance, but between two different types of learning, the deductive and inductive.

**Acknowledgments**

Our sincere and warmest thanks to all those who played the Interactive Minority Game. We are grateful to Damien Challet, Maya Paczuski and Duncan J. Watts for comments and advice, and especially Fabio Mariotti and Fribourg University Chemistry Department for providing us with a web server and much computational assistance. This work was supported by the Swiss National Science Foundation.

The Interactive Minority Game is online at

> http://www.unifr.ch/econophysics/minority/game/

**Table 1.** Statistics for the different markets of the Interactive Minority Game. Different market experiences are possible based on two parameters, MG agents' memory, *M*, and the number *N* of agents; the latter we have fixed (95 including the human) so that the market weight of the human is player is always the same. If the control parameter $\alpha = 2^M / N$ is below the critical value $\alpha_c \approx 0.3$, the market is in the "symmetric" phase; for $\alpha > \alpha_c$ the market phase is "asymmetric".

| Level | Agents' memory, *M* | $\alpha$ | Market phase | Volatility* $\sigma^2 / N$ | # of game turns played by humans § |
|---|---|---|---|---|---|
| Easy | 2 | 0.04 | symmetric | 7.58 | 15,400 |
| Apprentice | 3 | 0.08 | symmetric | 3.71 | 11,000 |
| Trader | 4 | 0.17 | symmetric | 1.46 | 9,400 |
| Professional | 6 | 0.67 | asymmetric | 0.25 | 9,800 |
| Guru | Mixed values | -- | asymmetric | 0.24 | 16,000 |

\* Data from simulations. $\sigma^2 / N = 1$ means that fluctuations are the same as if agents were playing randomly.

§ Data taken up until 05/2003. Note that this statistic only includes games of more than 100 turns in length.

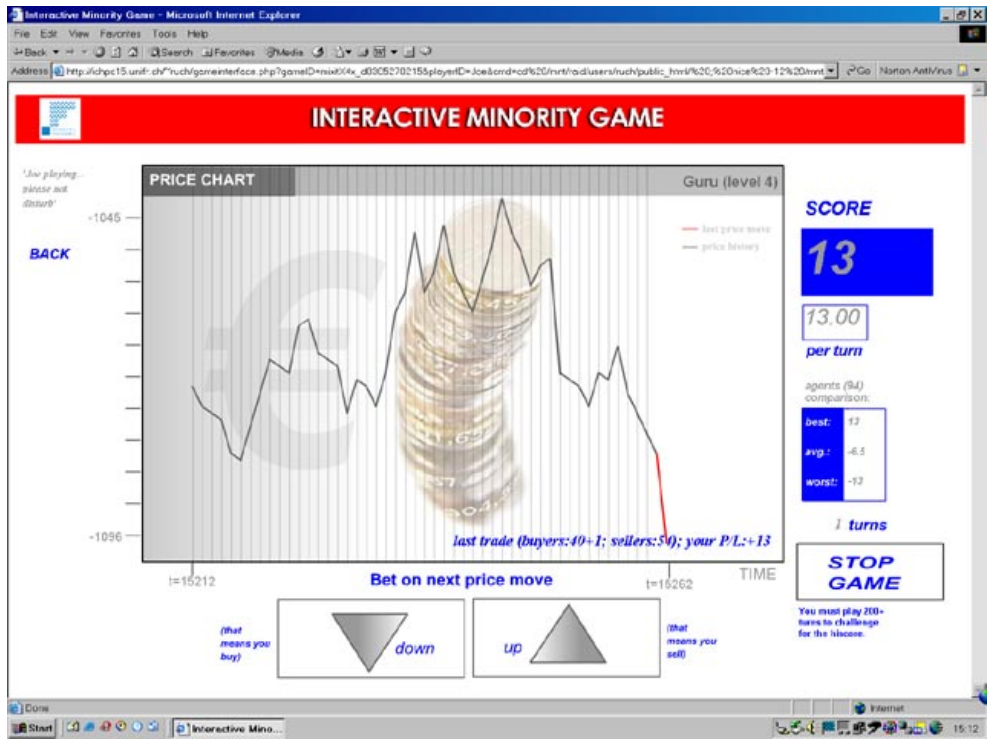

**Fig. 1.** The Interactive Minority Game online interface. The player is presented with a price history of the past 50 turns and must predict the direction of the next price movement.

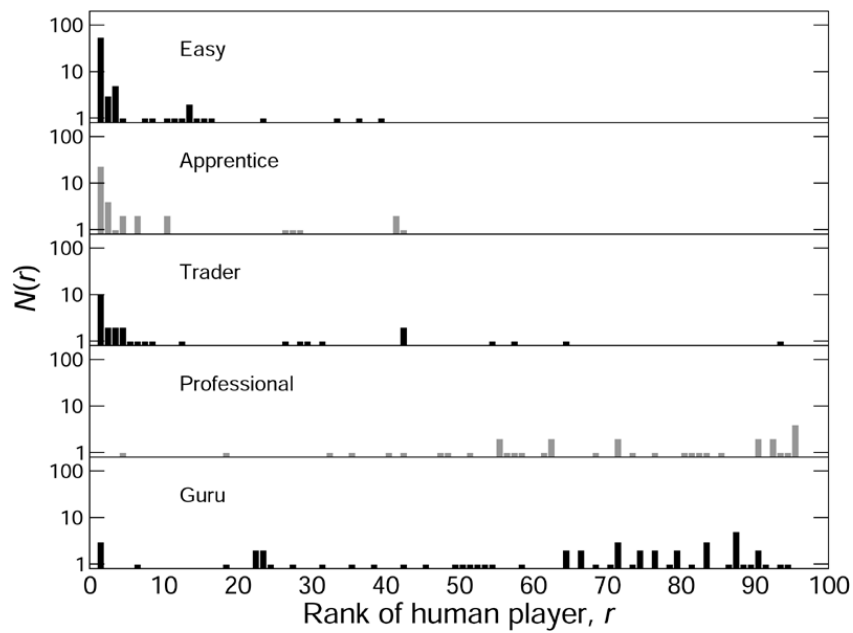

**Fig. 2.** Number $N(r)$ (log scale) of human players gaining rank $r$ among computer-controlled agents. $x$ and $y$ scales are the same for all graphs.

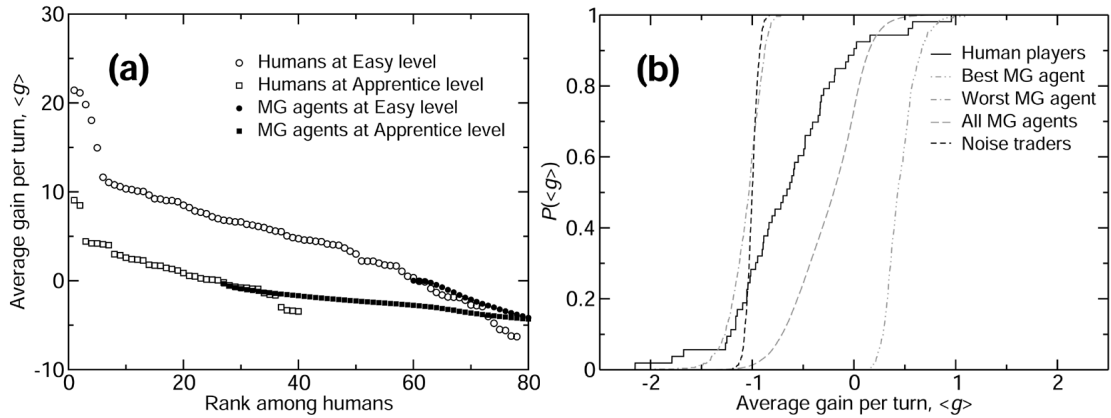

**Fig. 3.** *(a)* Score versus rank among other humans, at Easy (open circles) and Apprentice (open squares) levels, compared to scores of MG agents (filled symbols). In these easier markets, almost all human players are able to beat even the best MG agent. *(b)* Cumulative distribution of human scores at Guru level, compared to MG agents and noise trading. Simulation results averaged over 512 realisations; human results taken from all games of > 50 turns played up until 05/2003.

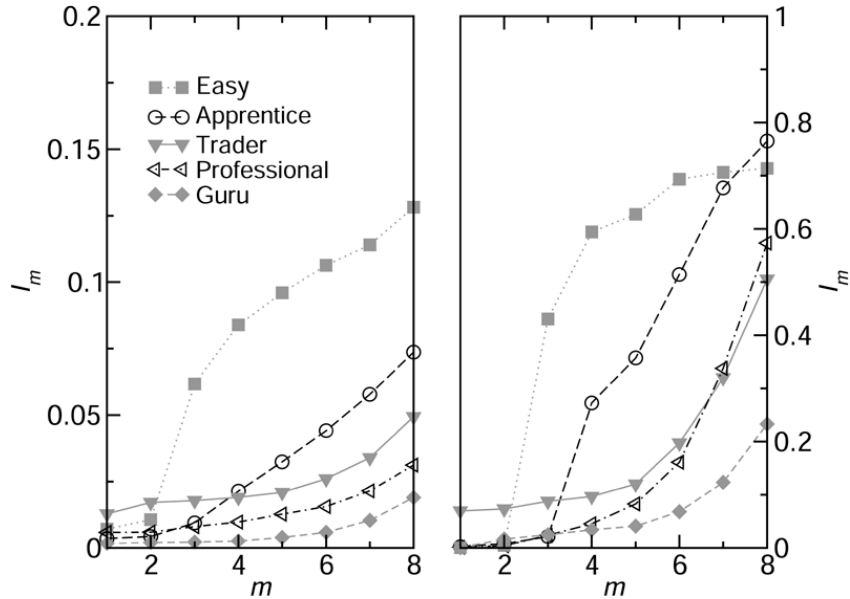

**Fig. 4.** Information gain ratio (Eq. 3) in human decisions conditional on market histories of length $m$. *Left:* Average over entire continuous market history. *Right:* Handpicked individual players with high scores. ($y$-axes are on different scales to allow easier view of data.) The discontinuity in $I_m$ at memory 3 and 4 respectively in the Easy and Apprentice levels indicates that human actions become predictable with respect to histories of length $m = M + 1$. The jump is larger for the handpicked players, indicating greater exploitation of market trends.

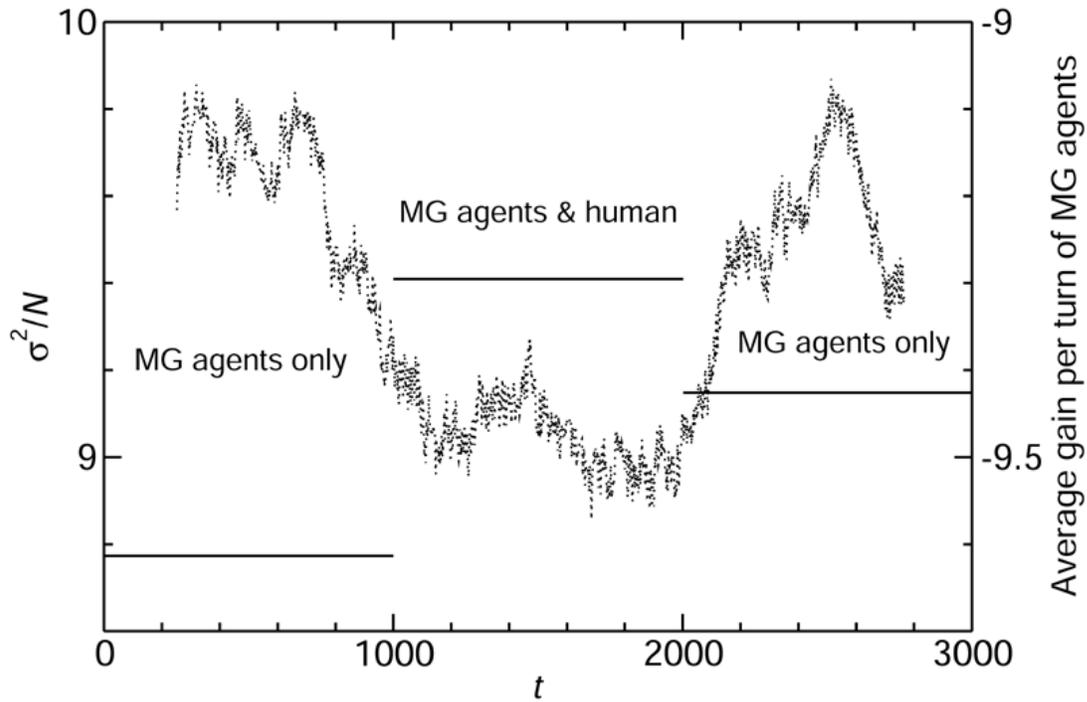

**Fig. 5.** Market volatility (Eq. 4) at Easy level with and without a human player (Peter Ruch). The first 1000 turns of the game are with MG agents only, followed by 1000 turns with a human player, and finally, 1000 again with MG agents only. The dotted line (left axis) plots the mean value of normalised volatility averaged over the 500-turn window ($t - 250$, $t + 250$). The presence of the human player decreases $\sigma^2/N$. The solid histogram shows MG agents' average gain per turn over the different periods (right axis), with an observable benefit being derived from the human speculator's presence.